\documentclass[conference]{IEEEtran}

\nonstopmode
\usepackage{amsmath,amssymb,amsfonts}
\usepackage{cite}
\usepackage{float} 
\usepackage{graphicx}
\usepackage{multicol}
\usepackage{url}
\usepackage{hyperref}

\usepackage{fancyhdr}
\newcommand\Transpose{^{\text{T}}}
\newcommand\bfa{  {\mathbf a} }
\newcommand\bfx{  {\mathbf x} }
\newcommand\bfy{  {\mathbf y} }
\newcommand\bfz{  {\mathbf z} }
\newcommand\calF{{\cal F}}
\newcommand\complete{^{\text{complete}}}
\newcommand\temp{}

\newcommand{\norm}{\survert}
\newcommand{\surbrace}[1]{ \left\{ #1 \right\} }
\newcommand{\surround}[1]{ \left(  #1 \right)  }
\newcommand{\survert}[1]{  \left\| #1 \right\| }

\newcommand\fraction[2]{\leavevmode\kern.1em\raise.5ex\hbox{\the\scriptfont0
#1}\kern-.1em/\kern-.15em\lower.25ex\hbox{\the\scriptfont0 #2}}

\newcommand\ZbeginE{ \begin{enumerate} }
\newcommand\ZendE{     \end{enumerate} }

\newcommand\ZbeginD{ \begin{description} }
\newcommand\ZendD{     \end{description} }

\newcommand{\SmallSpacingInLists}{%
\setlength{\topsep}{0pt}
\setlength{\parskip}{0pt}
\setlength{\partopsep}{0pt}
\setlength{\listparindent}{0pt}
\setlength{\parsep}{0pt}
\setlength{\itemsep}{0pt}}

\begin{document}
%---------------------------------------------------------------

\title{%
  \fbox{\parbox{2.4in}{\normalsize{\raggedright Originally presented in 2018 at NSSDF (Military Sensing Symposium) conference}}}
  \parbox{7in}{\ }
  \smallskip
  Game theory analysis when playing the wrong game} 
\author{\IEEEauthorblockN{Dan Zwillinger and Paul San Clemente}
\IEEEauthorblockA{\textit{BAE Systems Information and Electronic Systems Integration Inc.}\\
600 District Ave \\ 
Burlington, MA, 01803 \\
daniel.zwillinger@baesystems.com, paul.sanclemente@baesystems.com \\
}}
\maketitle

\begin{abstract}
  In classical game theory, optimal strategies are determined for
  games with complete information; this requires knowledge of the
  opponent's goals.  We analyze games when a player is mistaken about
  their opponents goals.  For definitiveness, we study the (common)
  bimatrix formulation where both player's payoffs are matrices.
  While the payoff matrix weights are arbitrary, we
  focus on strict ordinal payoff matrices, which can be enumerated.  In
  this case, a reasonable error would be for one player to switch two
  ordinal values in their opponents payoff matrix.  The mathematical
  formulation of this problem is stated, and all 78 strict ordinal
  2-by-2 bimatrix games are investigated.
  This type of incomplete information game has not -- to our knowledge -- been studied before.
\end{abstract}

\begin{IEEEkeywords}
game theory, optimal strategy, wrong game
\end{IEEEkeywords}

%---------------------------------------------------------------
\section{Introduction}
When playing a game it is important to know both your goals 
and your opponents goals.
If there is confusion about either of these, then a correct analysis
could result in the wrong strategy.

Brams \cite{Brams} gives an example from the 1979--1981 Iranian hostage
situation.
He analyzed the "game" that US President Carter was "playing" and
indicates how he (might have) arrived at his strategy.
Then Brams analyzed the different "game" that Khomeini was "playing."
Specifically, Bram modeled the Iranian hostage "game" using the
payoff matrices
\begin{equation*}
A=\begin{bmatrix}
4 & 2 \\
3 & 1 \\
\end{bmatrix}
\qquad
B_{\text{Carter}}=\begin{bmatrix}
3 & 4 \\
2 & 1 \\
\end{bmatrix}
\qquad
B_{\text{Khomeini}}=\begin{bmatrix}
2 & 4 \\
1 & 3 \\
\end{bmatrix}
\end{equation*}
where $A$ is the USA's payoff matrix, $B_{\text{Carter}}$ is what
Carter thought Khomeini's payoff matrix was and $B_{\text{Khomeini}}$
is what Khomeini's actual payoff matrix was.
Since the games were different, President Carter's strategy (derived
from analysis of $A$ and $B_{\text{Carter}}$) did not resolve the game
as he thought it would.

In game theory information can be "complete", which means that all
players know everything about the game (for example, Go, chess, and
checkers) or the game can be incomplete (for example, poker).
There are many types of incomplete information games.
This paper addresses one type of incomplete information game -- the
"cost" of playing a wrong game that is close to the correct game.
We believe this analysis is new.
We derive four problem statements representing different
interpretations of "playing the wrong game".
These cases represent different understandings of what the other
player thinks you are thinking.
We then systematically evaluate 
these cases for a class of interesting payoff matrices.

%---------------------------------------------------------------
\subsection{Assumptions and Terminology}

This paper considers two-player bimatrix games.
That is, player~1
has $n$ different strategies available for use,
uses strategy $\bfx\in\Delta$,
has payoff matrix $A$ of size $n\times m$,
and receives payoff $P_1$.
Here $\Delta$ is the simplex representing all possible (mixed and pure)
strategies\footnote{As usual, $\norm{\begin{bmatrix} c_1 &
      \dots & c_k\end{bmatrix}}_1=\sum_i^k |c_i|$.} $\Delta =
\surbrace{\bfz \mid 0\le z_i\le1, \norm{\bfz}_1=1}$.
Similarly, player~2 has the parameters $m$, $\bfy\in\Delta$, $B$,  and $P_2$.
The player payoffs are 
$P_1 = \bfx\Transpose A\bfy$ and
$P_2 = \bfx\Transpose B\bfy$.

We assume that the values in $A$ and $B$ are non-negative.
This means that the problem is \textit{not} zero sum; that is $A$ is not equal
to $-B$.

The simultaneous optimal strategies $\surbrace{\bfx^*,\bfy^*}$ are determined by
each player maximizing their payoff
\begin{equation}
  \begin{split}
  \bfx^* &= \arg\max_{\bfx\in\Delta}\bfx\Transpose A \bfy^*\\
  \bfy^* &= \arg\max_{\bfy\in\Delta}{\bfx^*}\Transpose B \bfy \\
  \end{split}
  \label{eq:base}
\end{equation}
where a superscript "T" represents vector transpose.
The solution of the general bimatrix game in (\ref{eq:base}) can be
found using the algorithm by Lemke and Howson~\cite{Lem64}.

%---------------------------------------------------------------
\section{Problem Statement}

This paper makes the following assumptions about the game being
played:
\ZbeginE \SmallSpacingInLists
\item Player~1 uses payoff matrix $A$; both players know this.
\item Player~1 believes that player~2 is using payoff matrix~$B$.
\item Matrix $B$ belongs to a family $\calF(B)$ of related matrices.
\item Player~2 is actually using payoff matrix $R$, which is in the family~$\calF(B)$.
\item The matrix $R$ may, or may not, be matrix~$B$.
\item Both payers are playing optimally, based on the information available to them.
  \ZendE
That is, player~1 has incomplete information; player~2's actual payoff
matrix is $R$, which player~1 does not know.

To precisely define the problem, we need to specify exactly what
the players believe to be true.
For player~1 we have two assumptions:
\ZbeginD \SmallSpacingInLists
\item[E\_] player~1 believes that player~2 is using the payoff matrix~$B$.
\item[F\_] player~1 believes that player~2 is using a payoff matrix in~$\calF(B)$.
\ZendD
where we use "E" to mean "exact"  and "F" to mean "family".
For player~2 we also have two assumptions:
\ZbeginD \SmallSpacingInLists
\item[\_E] player~2 believes that "player~1 believes that player~2 is using the payoff matrix~$B$."
\item[\_F] player~2 believes that "player~1 believes that player~2 is using a payoff matrix in~$\calF(B)$."
\ZendD
Combining these assumptions, there are four cases to analyze $\{EE, EF, FE, FF\}$.
Each case has different solutions, with the payoffs to player~1 being
$\surbrace{P_1^{EE}, P_1^{EF}, P_1^{FE}, P_1^{FF}}$.

A last case that could be studied is the "symmetric case", when each of the
two players has equal uncertainty about the other player's payoff matrix.
This game is described in section \ref{sec:sym}.

Note that if there were complete information, and player~1 knew that
player~2's payoff matrix was $R$, then the optimal strategies
$\{\bfx_R,\bfy_R\}$ would be found from (\ref{eq:base})
\begin{equation}
  \begin{split}
  \bfx_R &= \arg\max_{\bfx\in\Delta}\bfx\Transpose A \bfy_R \\
  \bfy_R &= \arg\max_{\bfy\in\Delta}\bfx_R\Transpose B \bfy \\
  \end{split}
  \label{eq:x:and:y:using:R}
\end{equation}
Hence, player~1's payoff would be
\begin{equation}
  P_1\complete = \bfx_R\Transpose A\bfy_R
  \label{eq:payoff:using:R}
\end{equation}
The purpose of this paper is to understand the difference between
$P_1\complete$ and $\surbrace{P_1^{EE}, P_1^{EF}, P_1^{FE},
  P_1^{FF}}$.

%---------------------------------------------------------------
\subsection{Computational Specification}

In this paper we evaluate solutions for all strict ordinal 2-by-2
bimatrix games.
An \textit{ordinal} payoff matrix contains relative integer preferences.
That is, a strategy that returns a "$k$" is preferred to a strategy
that returns a "$j$", if $k>j$.
A \textit{strict ordinal} payoff matrix of size $n$-by-$m$ contains the
entries $1,2,\dots,mn$ and each value is used exactly once; there are
no ties among the preferences.
Hence, all our payoff matrices contain each of the values $\{ 1, 2, 3,
4\}$ exactly once.
The Carter and Khomeini payoff matrices shown in the introduction are
strict ordinal 2-by-2 payoff matrices.

There are 78 unique strict ordinal 2-by-2 bimatrix games; see
Figure~\ref{fig:bimatrix}.
Uniqueness means that 2 matrix pairs cannot be made the same by
re-ordering strategies or switching players.
This value is derived in Rapoport and Guyer~\cite{Rap66}.
\begin{figure}
  \scriptsize
  \begin{multicols}{3}
    \ZbeginE
\item 1 3 2 4  -- 1 3 2 4
\item 4 1 2 3  -- 2 4 3 1
\item 4 1 2 3  -- 3 1 2 4
\item 4 1 2 3  -- 4 2 1 3
\item 4 1 3 2  -- 1 3 4 2
\item 4 1 3 2  -- 2 1 3 4
\item 4 1 3 2  -- 2 4 3 1
\item 4 1 3 2  -- 3 1 2 4
\item 4 1 3 2  -- 3 4 2 1
\item 4 1 3 2  -- 4 2 1 3
\item 4 1 3 2  -- 4 3 1 2
\item 4 2 1 3  -- 1 2 4 3
\item 4 2 1 3  -- 1 3 4 2
\item 4 2 1 3  -- 2 1 3 4
\item 4 2 1 3  -- 2 3 4 1
\item 4 2 1 3  -- 2 4 3 1
\item 4 2 1 3  -- 3 1 2 4
\item 4 2 1 3  -- 3 2 1 4
\item 4 2 1 3  -- 3 4 2 1
\item 4 2 1 3  -- 4 1 2 3
\item 4 2 1 3  -- 4 2 1 3
\item 4 2 1 3  -- 4 3 1 2
\item 4 2 3 1  -- 1 2 3 4
\item 4 2 3 1  -- 1 2 4 3
\item 4 2 3 1  -- 1 3 4 2
\item 4 2 3 1  -- 1 4 3 2
\item 4 2 3 1  -- 2 1 3 4
\item 4 2 3 1  -- 2 3 4 1
\item 4 2 3 1  -- 2 4 3 1
\item 4 2 3 1  -- 3 1 2 4
\item 4 2 3 1  -- 3 2 1 4
\item 4 2 3 1  -- 3 4 1 2
\item 4 2 3 1  -- 3 4 2 1
\item 4 2 3 1  -- 4 1 2 3
\item 4 2 3 1  -- 4 2 1 3
\item 4 2 3 1  -- 4 3 1 2
\item 4 2 3 1  -- 4 3 2 1
\item 4 3 1 2  -- 1 2 3 4
\item 4 3 1 2  -- 1 2 4 3
\item 4 3 1 2  -- 1 3 4 2
\item 4 3 1 2  -- 1 4 3 2
\item 4 3 1 2  -- 2 1 3 4
\item 4 3 1 2  -- 2 1 4 3
\item 4 3 1 2  -- 2 3 1 4
\item 4 3 1 2  -- 2 3 4 1
\item 4 3 1 2  -- 2 4 3 1
\item 4 3 1 2  -- 3 1 2 4
\item 4 3 1 2  -- 3 2 1 4
\item 4 3 1 2  -- 3 2 4 1
\item 4 3 1 2  -- 3 4 1 2
\item 4 3 1 2  -- 3 4 2 1
\item 4 3 1 2  -- 4 1 2 3
\item 4 3 1 2  -- 4 1 3 2
\item 4 3 1 2  -- 4 2 1 3
\item 4 3 1 2  -- 4 3 1 2
\item 4 3 1 2  -- 4 3 2 1
\item 4 3 2 1  -- 1 2 3 4
\item 4 3 2 1  -- 1 2 4 3
\item 4 3 2 1  -- 1 3 2 4
\item 4 3 2 1  -- 1 3 4 2
\item 4 3 2 1  -- 1 4 2 3
\item 4 3 2 1  -- 1 4 3 2
\item 4 3 2 1  -- 2 1 3 4
\item 4 3 2 1  -- 2 1 4 3
\item 4 3 2 1  -- 2 3 1 4
\item 4 3 2 1  -- 2 3 4 1
\item 4 3 2 1  -- 2 4 3 1
\item 4 3 2 1  -- 3 1 2 4
\item 4 3 2 1  -- 3 1 4 2
\item 4 3 2 1  -- 3 2 1 4
\item 4 3 2 1  -- 3 2 4 1
\item 4 3 2 1  -- 3 4 1 2
\item 4 3 2 1  -- 3 4 2 1
\item 4 3 2 1  -- 4 1 2 3
\item 4 3 2 1  -- 4 1 3 2
\item 4 3 2 1  -- 4 2 1 3
\item 4 3 2 1  -- 4 2 3 1
\item 4 3 2 1  -- 4 3 1 2
\ZendE
\end{multicols}
  \caption{All 78 strict ordinal 2-by-2 bimatrix games; the first 4
    elements are $A$ (column-wise), the next 4 elements are $B$
    (column-wise).}
  \label{fig:bimatrix}
\end{figure}

For a payoff matrix $B$, we define the family $\calF(B)$ to contain 4
matrices: $B$ and the 3 matrices derivable from $B$ by switching a
pair of adjacent preferences.
(These are called "swaps" by \cite{Rob05}.)
That is, given $B$ the matrix $B_{12}$ is obtained by switching the values "1" and "2",
the matrix $B_{23}$ is obtained by switching the values "2" and "3", and 
the matrix $B_{34}$ is obtained by switching the values "3" and "4".
The motivation for this type of incomplete information is that we may
generally understand someone else's preferences, but we may make a
single mistake.
For example, suppose you articulated your best friend's top 4 dessert choices.
Are you confident that you have not switched their 2nd and 3rd preferences?

Here is a numerical example: the payoff matrix $B = \begin{bmatrix} 1
  & 3 \\ 4 & 2 \\ \end{bmatrix}$ is in the same family as
\begin{equation*}
B_{12} = \begin{bmatrix}  \bf{2} & 3 \\ 4 & \bf{1} \\ \end{bmatrix}  \quad
B_{23} = \begin{bmatrix}  1 & \bf{2} \\ 4 & \bf{3} \\ \end{bmatrix}  \quad
B_{34} = \begin{bmatrix}  1 & \bf{4} \\ \bf{3} & 2 \\ \end{bmatrix}
\end{equation*}

%---------------------------------------------------------------
\section{Problem Statement -- Mathematically}

This section contains mathematical descriptions of the different games of interest.

%---------------------------------------------------------------
\subsection{Game EE}
In this case
\ZbeginD \SmallSpacingInLists
\item[E\_] player~1 believes that player~2 is using the payoff matrix~$B$.
\item[\_E] player~2 believes that "player~1 believes that player~2 is using the payoff matrix~$B$."
\ZendD
From assumption "E\_" player~1 solves the system (from
(\ref{eq:base})) for the strategies $\surbrace{\bfx_{EE},\bfy_*}$:
\begin{equation}
  \begin{split}
  \bfx_{EE} &= \arg\max_{\bfx\in\Delta}\bfx\Transpose A \bfy_*  \\
  \bfy_* &= \arg\max_{\bfy\in\Delta}\bfx_{EE}\Transpose B \bfy \\
  \end{split}
  \label{eq:x:EE}
\end{equation}
From assumption "\_E" player~2 knows that player~1 will solve 
(\ref{eq:x:EE}), and is using the strategy $\bfx_{EE}$.
However, since player~2 is using payoff matrix $R$, player~2
determines their strategy from
\begin{equation}
  \bfy_{EE} = \arg\max_{\bfy\in\Delta}\ \bfx_{EE}\Transpose R \bfy
  \label{eq:EE:next}
\end{equation}
This can be readily solved by determining the indices $I$ of the
maximum values of the vector $\bfx_{EE}\Transpose R$; all other
indices in $\bfy_{EE}$ have the numerical value zero (that is, $y_i=0$ for
$i\not\in I$).
Any values can be used for $y_i$ when $i\in I$, consistent with
$\bfy\in\Delta$; the payoff to player~2 will be the same.

%---------------------------------------------------------------
\subsection{Game FE}
In this case
\ZbeginD \SmallSpacingInLists
\item[F\_] player~1 believes that player~2 is using a payoff matrix in~$\calF(B)$.
\item[\_E] player~2 believes that "player~1 believes that player~2 is using the payoff matrix~$B$."
\ZendD
That is, player~2 is confused about what player~1 knows.
By "\_E" player~2 thinks that player~1 will perform the same
analysis that player~1 performed in (\ref{eq:x:EE}).
Then as before, player~2 calculates the quantity in
(\ref{eq:EE:next}), re-written here is new variables and explicitly
showing the dependence on $R$:
\begin{equation}
  \bfy_{FE}(R) = \arg\max_{\bfy\in\Delta} \ \bfx_{EE}\Transpose R \bfy
  \label{eq:FE:next}
\end{equation}
However, by "F\_", player~1 knows not to use $\bfx_{EE}$ but needs to incorporate $\calF$ thinking.
That is, player~1 needs to solve
\begin{equation}
  \bfx_{FE} = \arg \max_{\bfx} \surround{ \min_{R\in \calF(B)} \bfx\Transpose A \bfy_{FE}(R) }
  \label{eq:FE:x}
\end{equation}
to ensure a maximal return, regardless of which $R\in \calF(B)(B)$ that player~2 chooses.

This can be solved as follows.
First, define $\bfa_j = A \bfy_{EF}(R_j)$ for each $R_j\in \calF(B)$.
Then\footnote{The authors thank Peter Kingston for this observation.}
we note that (\ref{eq:FE:x}) can be written as
\begin{equation}
  \begin{split}
    \bfx_{FE} &= \arg \max_{\bfx} \surround{ \min_{R_j\in \calF} \bfx\Transpose \bfa_j }\\
             &= \arg \max_{\bfx} \surround{ \max_{z} \surbrace{ z \mid z \le \bfx\Transpose \bfa_j} }\\
  \end{split}
\end{equation}
This, in turn, can be written as a linear programming problem for the
values of $\bfz$ and $z$:
\begin{equation}
  \begin{split}
    \max_{\{\bfx,z\}} \qquad & z \\
     z &\le \bfa_j\Transpose \bfx \qquad \text{for all $j$}\\
      \norm{\bfx}_1 &= 1\\
            0 &\le x_i \le 1 \\
            0 &\le z \\
  \end{split}
  \label{eq:lp}
\end{equation}
The $\bfx$ part of the solution of (\ref{eq:lp}) is now~$\bfx_{FE}$.

%---------------------------------------------------------------
\subsection{Game FF}
\label{sec:FF}
In this case
\ZbeginD \SmallSpacingInLists
\item[F\_] player~1 believes that player~2 is using a payoff matrix in~$\calF(B)$.
\item[\_F] player~2 believes that "player~1 believes that player~2 is using a payoff matrix in~$\calF(B)$."
\ZendD
Here, player~1 does not know player~2's payoff matrix, but knows the
family ($\calF$) that contains it.
Hence, player~1 needs to solve the following to find $\bfx_{FF}$:
\begin{equation}
  \begin{split}
    \bfx_{FF}   &= \arg \max_{\bfx} \surround{  \min_{R\in \calF(B)} \bfx    \Transpose A \bfy_{*}(R)} \\
    \bfy_{*}(R) &= \arg \max_{\bfy} \                   \bfx_{FF}\Transpose R \bfy     \\
  \end{split}
  \label{eq:x:FF}
\end{equation}
After player~1 obtains the optimal $\bfx_{FF}$, player~2 will solve
the same problem as in (\ref{eq:FE:next}), rewritten in FF variables:
\begin{equation}
  \bfy_{FF} = \arg\max_{\bfy\in\Delta} \ \bfx_{FF}\Transpose R \bfy
  \label{eq:y:FF}
\end{equation}
It is possible to determine the optimal strategy ($\bfx_{FF}$) for
player~1, as defined by (\ref{eq:x:FF}), using advanced game theory
techniques.
For example, we could represent the problem in sequence form and
then solve a complementary problem, see \cite[Theorem 3.14]{Nisan}.

Rather than elaborate on that solution technique, let's refer to a
solution of (\ref{eq:x:FF}) as $\bfx_{FF} = M(A,\calF)$.
In section \ref{sec:M:function}, we show how to determine $M$ for the
payoff matrices that are of interest to us.

%---------------------------------------------------------------
\subsection{Game EF}
In this case
\ZbeginD \SmallSpacingInLists
\item[E\_] player~1 believes that player~2 is using the exactly the payoff matrix~$B$.
\item[\_F] player~2 believes that "player~1 believes that player~2 is using a payoff matrix in~$\calF(B)$."
\ZendD

From player~1's point of view, Game EF is no different from Game EE.
Hence, the solution ($\bfx_{EF}$) is the same as the solution in
(\ref{eq:x:EE}); $\bfx_{EF}=\bfx_{EE}$.
  Writing those equations in EF variables:
\begin{equation}
  \begin{split}
  \bfx_{EF} &= \arg\max_{\bfx\in\Delta}\bfx\Transpose A \bfy_*  \\
  \bfy_* &= \arg\max_{\bfy\in\Delta}\bfx_{EF}\Transpose B \bfy \\
  \end{split}
  \label{eq:x:EF}
\end{equation}

From player~2's point of view, game EF is no different from Game FF.
Hence, the solution ($\bfy_{EF}$) is the same as the solution in
(\ref{eq:x:FF}) and (\ref{eq:y:FF}).
  Writing those equations in EF variables:
\begin{equation}
  \begin{split}
    \bfx_{EF} &= M(A,\calF) \\
    \bfy_{EF} &= \arg\max_{\bfy\in\Delta} \ \bfx_{EF}\Transpose R \bfy \\
  \end{split}
  \label{eq:x:and:y:EF}
\end{equation}

%---------------------------------------------------------
\subsection{The symmetric game}
\label{sec:sym}
%---------------------------------------------------------
In the symmetric game each player has the same knowledge; each knows
that the other player is using a payoff matrix within a known family
of payoffs.
Specifically, player~1 knows
\ZbeginE % \SmallSpacingInLists
\item player~1 uses known payoff matrix $R_A\in\calF(A)$
\item player~2 uses an unknown payoff matrix in $\calF(B)$
\ZendE
while, player~2 knows (symmetrically)
\ZbeginE % \SmallSpacingInLists
\item player~2 uses a known payoff matrix $R_B\in\calF(B)$
\item player~1 uses an unknown payoff matrix in $\calF(A)$
\ZendE
Assume that players 1 and 2 play the strategies $\bfx_S$ and $\bfy_S$.
Then, player~1 knows that player~2 is going to play one of the strategies
\begin{equation}
  \bfy_{S}^i(\bfx_S) = \arg \max_{\bfy} \ \bfx_S\Transpose R_E^i \bfy
  \qquad  R_E^i\in \calF(B)
  \label{eq:sym:1a}
\end{equation}
for $i=1,...,4$.
Player~1 wants to maximize his payoff, so his strategy is given by
\begin{equation}
  \bfx_{S} = \arg \max_{\bfx} \surround{ \min_i \  \bfx\Transpose R_A \bfy_S^i(\bfx)}
  \label{eq:sym:1b}
\end{equation}
That is, player~1's strategy only depends on the matrices $A$ and $B$
and is given by equations (\ref{eq:sym:1a}) and (\ref{eq:sym:1b}).
By symmetry, layer 2's strategy will be
\begin{equation}
  \begin{split}
  \bfx_{S}^j(\bfy) &= \arg \max_{\bfx} \ \bfx\Transpose R_A^j \bfy
  \qquad  R_A^j\in \calF(A) \\
  \bfy_{S} &= \arg \max_{\bfy}  \surround{\min_j \ \bfx_S^j(\bfy)}\Transpose R_E \bfy \\
  \end{split}
  \label{eq:sym:2}
\end{equation}

Once again, this game can be solved by using the extensive game form and complementarity.
This variation is not considered further in this paper.

%---------------------------------------------------------
\subsection{The function $M(A,\calF)$}
\label{sec:M:function}
%---------------------------------------------------------
%
The examples in this paper have payoff matrices of size 2-by-2.
For problems of this size, we can readily evaluate the $M$ function
defined in section \ref{sec:FF}.
The following example shows how it can be done.

Consider the payoff matrices $A$ and $B$:
\begin{equation}
A = \begin{bmatrix}  4 & 1 \\ 2 & 3 \\ \end{bmatrix}
\qquad
B = \begin{bmatrix}  1 & 3 \\ 4 & 2 \\ \end{bmatrix}
\label{eq:example}
\end{equation}
The family that contains $B$ is $\calF(B)=\surbrace{B,B_{12},B_{23},B_{34}}$ with 
\begin{equation*}
B_{12} = \begin{bmatrix}  2 & 3 \\ 4 & 1 \\ \end{bmatrix}  \quad
B_{23} = \begin{bmatrix}  1 & 2 \\ 4 & 3 \\ \end{bmatrix}  \quad
B_{34} = \begin{bmatrix}  1 & 4 \\ 3 & 2 \\ \end{bmatrix}
\end{equation*}
Since the payoff matrices are two-dimensional, the strategies ($\bfx$
and $\bfy$) are also two-dimensional.
Therefore we can, without loss of generality,
write each of the vectors $\bfx$ and $\bfy$ in terms of one free parameter
\begin{equation*}
  \begin{split}
    \bfx &= \begin{bmatrix} a & 1-a \end{bmatrix}\Transpose \\
    \bfy &= \begin{bmatrix} b & 1-b \end{bmatrix}\Transpose \\
  \end{split}
  \label{eq:a:b:vecs}
\end{equation*}
where $0\le a\le1$ and $0\le b\le1$.
Then, by explicit computation, we can determine the payoffs to each player.
Using $P_N^C$ to denote the payoff to player $N$ who has payoff matrix $C$:
\begin{equation*}
  \begin{split}
    P_1^A       &=  \bfx\Transpose A     \bfy = (4b-2)a+(3-b)  \\
    P_2^{B}     &=  \bfx\Transpose B    \bfy = (2-4a)b+(2+a)\\
    P_2^{B_{12}} &=  \bfx\Transpose B_{12} \bfy = (3-4a)b+(1+2a)\\
    P_2^{B_{23}} &=  \bfx\Transpose B_{23} \bfy = (1-2a)b+(3-a)\\
    P_2^{B_{34}} &=  \bfx\Transpose B_{34} \bfy = (1-4a)b+(2+2a)\\
    &   \\
  \end{split}
\end{equation*}
Consider the different payoffs to player~2; each has the form
$()_1b+()_2$; where the parenthesis terms do not contain a~$b$.
Hence, for player~2 to maximize their payoff, we have:
\begin{equation*}
  \begin{split}
    &\text{if $()_1>0$ then $b=1$}   \\
    &\text{if $()_1<0$ then $b=0$}   \\
  \end{split}
\end{equation*}
and the value of $b$ is undefined otherwise.
Using this logic, we find the following choices for player~2:
\begin{equation*}
  \begin{split}
    \text{for  $B$ player~2 plays} &\begin{cases}
      b=1 & \text{if $a<\fraction{1}{2}$} \\
      b=0 & \text{if $a>\fraction{1}{2}$} \\
      \end{cases} \\
    \text{for  $B_{12}$ player~2 plays} &\begin{cases}
      b=1 & \text{if $a<\fraction{3}{4}$} \\
      b=0 & \text{if $a>\fraction{3}{4}$} \\
      \end{cases} \\
    \text{for  $B_{23}$ player~2 plays} &\begin{cases}
      b=1 & \text{if $a<\fraction{1}{2}$} \\
      b=0 & \text{if $a>\fraction{1}{2}$} \\
      \end{cases} \\
    \text{for  $B_{34}$ player~2 plays} &\begin{cases}
      b=1 & \text{if $a<\fraction{1}{4}$} \\
      b=0 & \text{if $a>\fraction{1}{4}$} \\
      \end{cases} \\
    &   \\
  \end{split}
\end{equation*}

\begin{figure}
  \includegraphics[width=\linewidth]{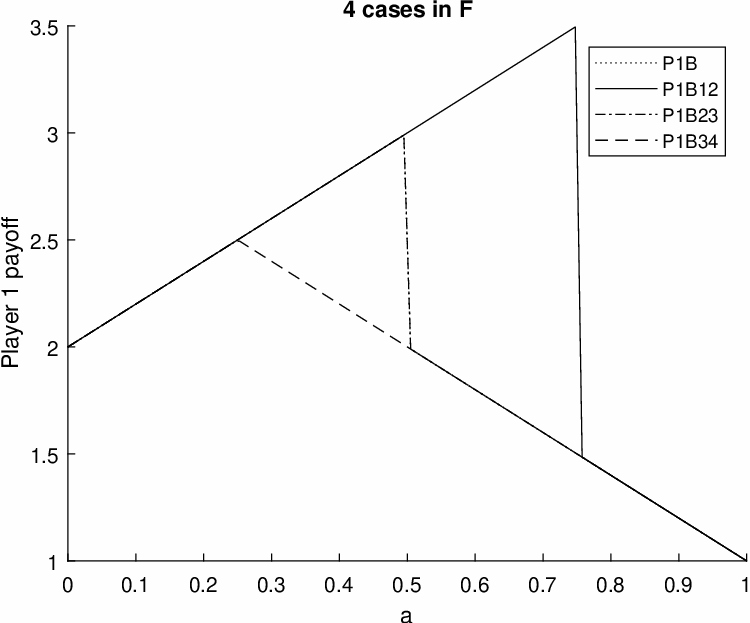}
  \caption{The payoffs to player~1 based on different player~2 payoff matrices; using (\ref{eq:example}).}
  \label{fig:P1}
\end{figure}

Using these values, we can determine the payoff to player~1 (that
is, $P_1^A$) using the different payoff matrices in~$\calF(B)$; see Figure~\ref{fig:P1}.
In this case, the maximum value of the payoff to player~1, minimized
over all matrices in $\calF(B)$ is given by $a=\fraction14$ for which $P_1(a)=2.5$.

With 2-by-2 payoff matrices, the payoffs will always be bilinear in
$a$ and $b$; that is, the payoff has the form $c_0 +
c_aa+c_bb+c_{ab}ab$ for constants $\{c_0,c_a,c_b,c_{ab}\}$.
Hence, the value of $b$ (determined by player~2 to maximize their
payoff) will always be zero or one, on different sides of an $a$
\textit{breakpoint}.

Then, for each player~2 payoff matrix, player~1's payoff is composed
of two different linear functions of $a$, the change between the two
occurring when $b$ changes between zero and one; which corresponds to
an $a$ breakpoint.
Hence, Figure~\ref{fig:P1} is representative of many cases (except,
sometimes, the linear functions will have slopes with different signs).

\begin{figure}
  \includegraphics[width=\linewidth]{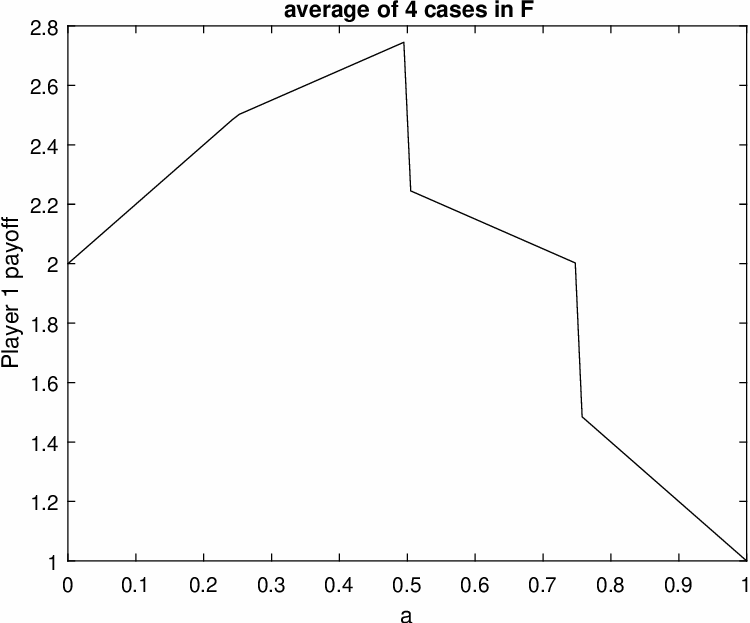}
  \caption{The payoffs to player~1, averaging the different player~2 payoff matrices; using (\ref{eq:example}).}
  \label{fig:P1:average}
\end{figure}

Alternatively, if player~1 knew that all of the matrices in the family
$\calF(B)$ were equally likely (i.e., each has a probability of
one-fourth), then the maximizing solution for player~1 would be to
determine the best payoff using the \textit{average} value of the
payoffs, see Figure~\ref{fig:P1:average}.
In this case, player~1 would select $a=0.5^{-}$ with an average payoff of
$P_1(a) = 2.75$.  This improves the payoff to player~1 by 10\%.
The notion of average payoff will not be pursued further in this
paper.

\begin{figure}
  \includegraphics[width=\linewidth]{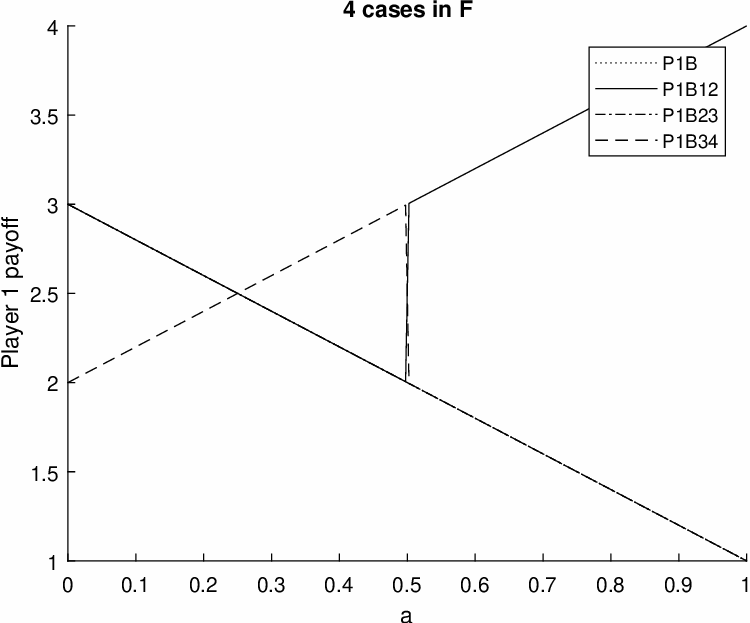}
  \caption{The payoffs to player~1 based on different player
    2 payoff matrices; using (\ref{eq:A4123:B1234}).}
  \label{fig:P1:alt}
\end{figure}

\begin{figure}
  \includegraphics[width=\linewidth]{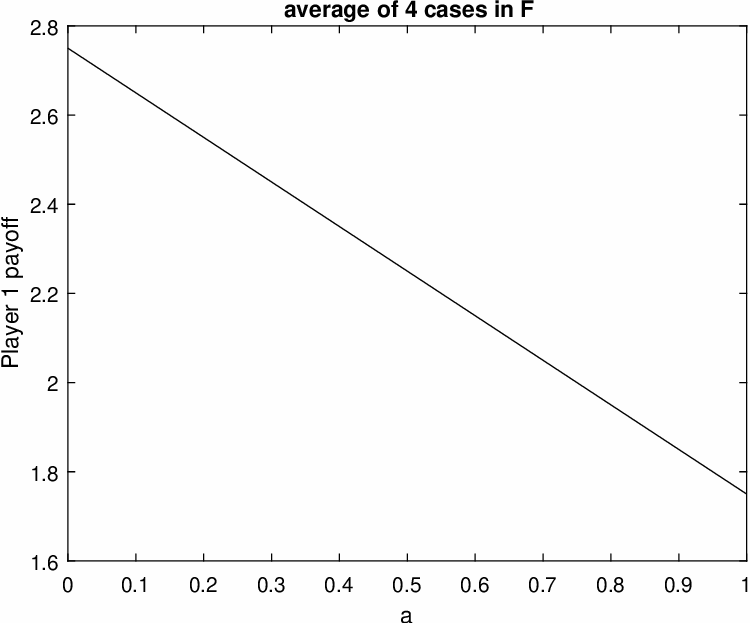}
  \caption{The payoffs to player~1, averaging the different
    player~2 payoff matrices; using (\ref{eq:A4123:B1234}).}
  \label{fig:P1:average:alt}
\end{figure}

Another example, which shows that the optimal value of $a$ does not
need to occur at the breakpoint for a single payoff matrix is given by
the matrices
\begin{equation}
A = \begin{bmatrix}  4 & 1 \\ 2 & 3 \\ \end{bmatrix}
\qquad
B = \begin{bmatrix}  1 & 2 \\ 3 & 4 \\ \end{bmatrix}
\label{eq:A4123:B1234}
\end{equation}
In this case, the results corresponding to graphs (\ref{fig:P1}) and
(\ref{fig:P1:average}) are in graphs (\ref{fig:P1:alt}) and
  (\ref{fig:P1:average:alt}).

\goodbreak
%---------------------------------------------------------
\section{Numerical Results for strict ordinal games}
%---------------------------------------------------------
%
The numerical results in this paper were obtained using Matlab by the following process:
\ZbeginE % \SmallSpacingInLists
\item Consider every bimatrix pair in the class of strict ordinal 2-by-2 payoffs
\item For each, determine the solutions $\{\bfx,\bfy\}$ for the four cases $\{EE,EF,FE,FF\}$
\item For each solution, determine the payoff to player~1 and compare
  it to the complete information payoff in (\ref{eq:payoff:using:R}).

  \ZendE
  A summary of the numerical results is in the following table

\renewcommand\temp[3]{ #1 & #2 & #3 \\ \hline}

\begin{center}
  \begin{tabular}{|c|c|c|}
    \hline
    \temp{Type}{Number of games with no change}{average loss}
    \temp{EE}{44}{1.8676}
    \temp{EF}{56}{1.7045}
    \temp{FE}{44}{1.3971}
    \temp{FF}{44}{0.9632}
  \end{tabular}
\end{center}
These values should be compared to the average payoff to player~1 in
the complete information game in (\ref{eq:base}), which is 3.481.
For example, in the ``EE'' case, player~1 has (on average) lost 54\% of
the value that could have been obtained.

Below are graphs showing the numerical results; both the absolute and
relative change in payoff to Player 1 under the four variations
considered.
The results are shown only for those games for which the payoff to player~1 changes.

%---------------------------------------------------------
% \subsection{Game EE analysis}
%---------------------------------------------------------
\begin{figure}[H]
  \includegraphics[width=1.6in]{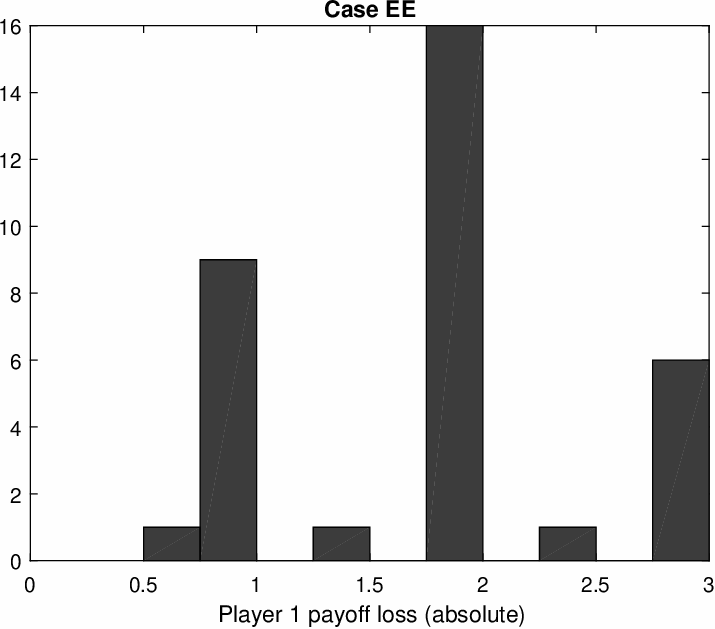}
  \quad
  \includegraphics[width=1.6in]{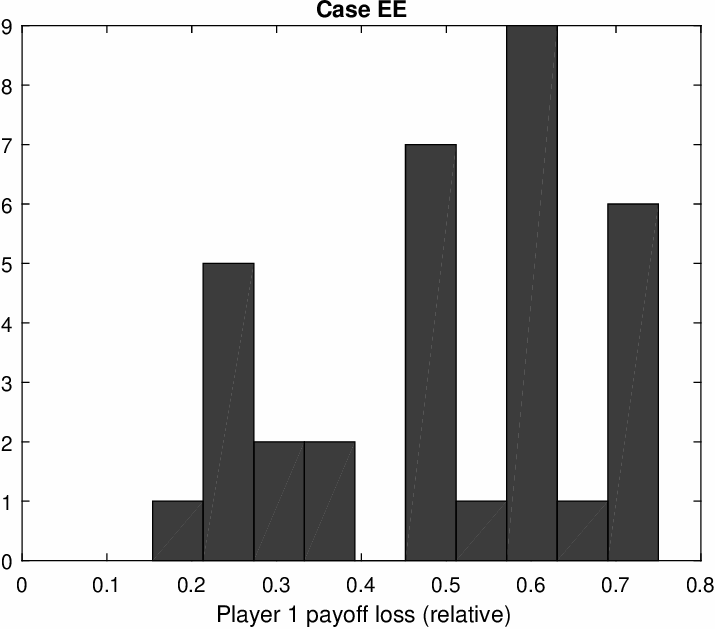}
  \caption{EE histogram: absolute (left) and relative (right) loss.}
  \label{fig:result:EE}
\end{figure}

%---------------------------------------------------------
% \subsection{Game EF analysis}
%---------------------------------------------------------
\begin{figure}[H]
  \includegraphics[width=1.6in]{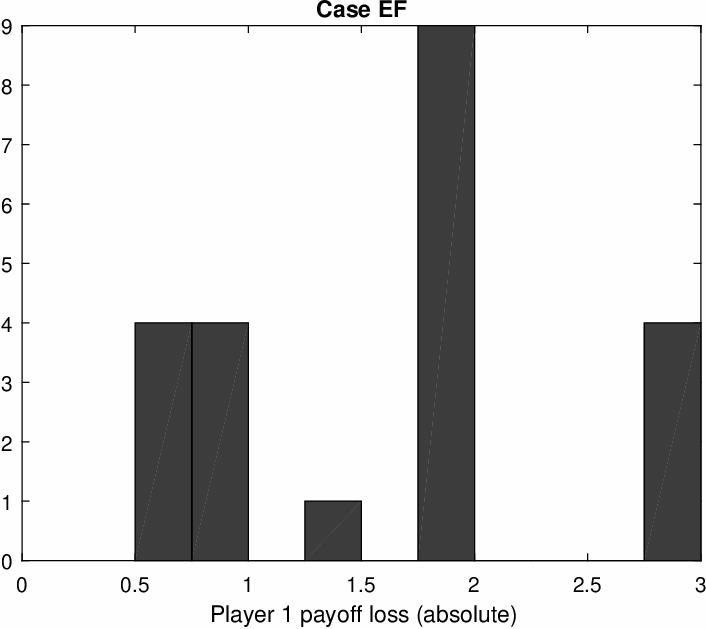}
  \quad
  \includegraphics[width=1.6in]{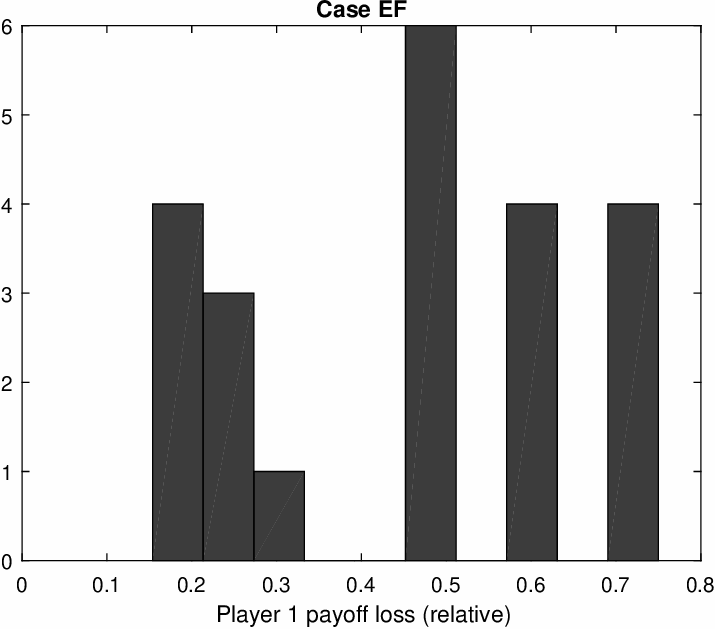}
  \caption{EF histogram: absolute (left) and relative (right) loss.}
  \label{fig:result:EF}
\end{figure}

%---------------------------------------------------------
% \subsection{Game FE analysis}
%---------------------------------------------------------
\begin{figure}[H]
  \includegraphics[width=1.6in]{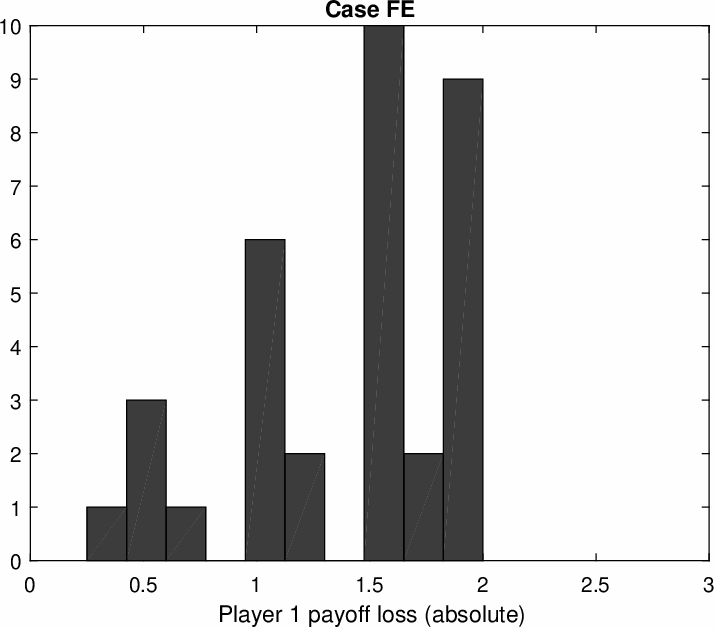}
  \quad
  \includegraphics[width=1.6in]{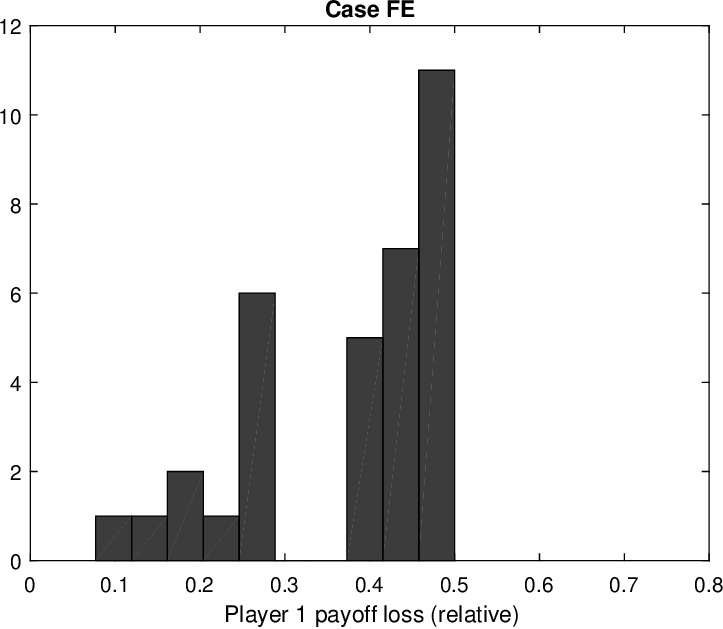}
  \caption{FE histogram: absolute (left) and relative (right) loss.}
  \label{fig:result:FE}
\end{figure}

%---------------------------------------------------------
% \subsection{Game FF analysis}
%---------------------------------------------------------
\begin{figure}[H]
  \includegraphics[width=1.6in]{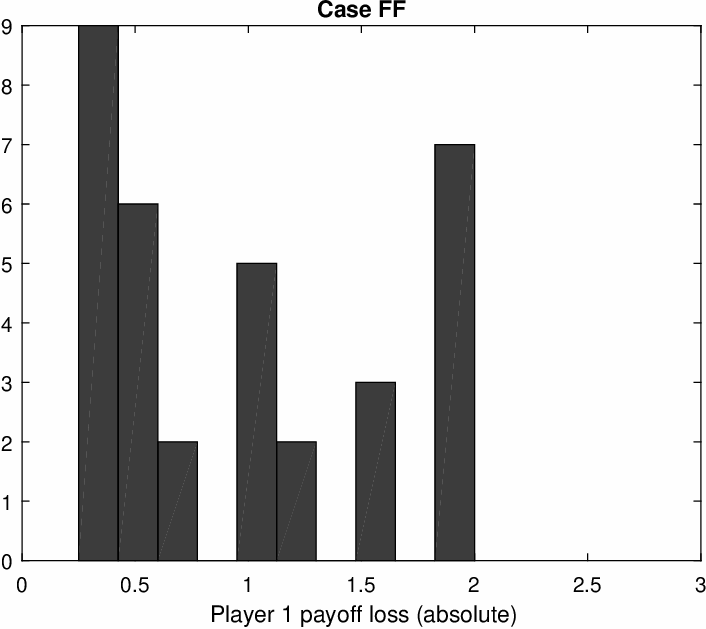}
  \quad
  \includegraphics[width=1.6in]{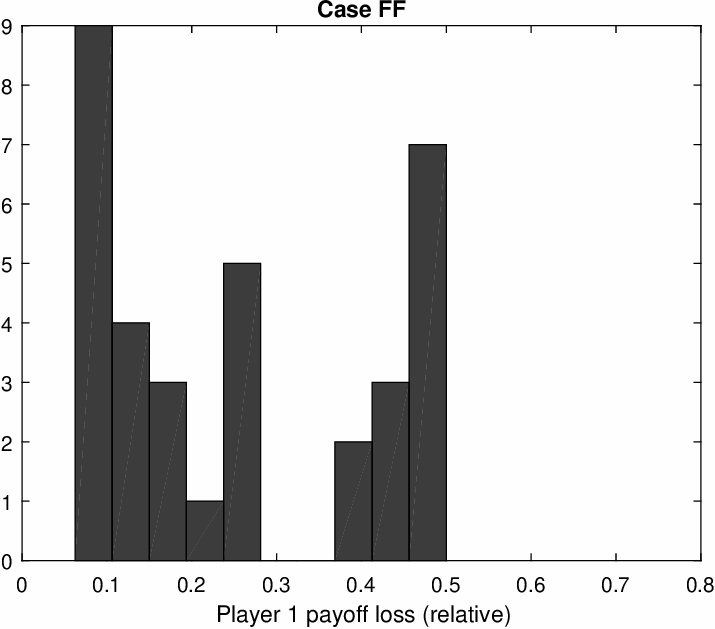}
  \caption{FF histogram: absolute (left) and relative (right) loss.}
  \label{fig:result:FF}
\end{figure}

%---------------------------------------------------------
\section{Observations and Conclusion}
%---------------------------------------------------------
We note the following:
\ZbeginE % \SmallSpacingInLists

\item Under the all variations considered, in the average case,
  player~1's payoff is reduced compared to the perfect
  information case.

\item In every case, adding additional uncertainty to either player~1
  or player~2 (by transitioning from "E" to "F") results in a worse
  payoff for player~1 (on average over all 78 games).  However, in the
  ``FE'' case, there are specific games for which player~1 obtains a
  small improvement.
  
\item For 22 games the payoffs to player~1 do not change under any of
  the game variations considered.
  Similarly, in 44 games, the payoffs to player~1 are worse in every game variation.
  For 12 games the payoff to player~1 is worse, but only in the ``EF'' case.

\item
  In the games "EE" and "EF" the maximum loss to player~1 can be
  as large as 3 (which is the maximal loss, going from 4 to 1).
  In the games "FE" and "FF" the maximum loss to player~1 is only~2.  
  
\item
  There are several "famous" two person games, here is how they relate to our game analysis:
\ZbeginE % \SmallSpacingInLists
\item no change in payoff to player~1 under any of the four variations:
"chicken",
"hero",
  "no conflict",
  and
"stag hunt"
  
\item reduced payoff to player~1 under every one of the four variations: 
"assurance",
"battle of the sexes",
"compromise",
"coordination game",
"deadlock",
"peace",
  and
  "prisoner's dilemma"

\item reduced payoff only in the ``EF''  variation: "harmony"

\ZendE

\ZendE

In all cases, from the perspective of player~1's average payoff, the
games ``EE'' and ``EF'' are worse than the games ``FE'' and ``FF''.

%---------------------------------------------------------
% \subsection{Potential future work}
%---------------------------------------------------------
%
Future work that would extend the analysis in this paper:
\ZbeginE \SmallSpacingInLists

\item A complete analysis of the symmetric game could be performed.

\item An evaluation could be made of how the payoff to player~2
  changes under the variations considered.  This would allow questions
  such as the following to be answered: ``How valuable is it to
  Player~2 to change what Player 1 is thinking about Player~2?''

\item
  As we've noticed, the payoff to Player 1 does not change under many
  games; these are games for which it is not in Player 1's interest to
  learn more about the Player 2's payoff.  Understanding the feaures
  of these games would be useful, especially if the information could
  be extrapolated to larger non-ordinal games.
  
  \ZendE

%---------------------------------------------------------
% \section*{References}
%---------------------------------------------------------

\bibliographystyle{./IEEEtran}
\bibliography{./IEEEabrv,./bibliography}

\begin{thebibliography}{9}

\bibitem[Bra93]{Brams}
  Steven J. Brams,
  "Theory of Moves",
  \textit{American Scientist},
  November--December 1993,
  pp 562--570,
  \url{https://www.acsu.buffalo.edu/~fczagare/Game%20Theory/Theory%20of%20Moves.pdf}
    (accessed 29 May 2018).
        
\bibitem[Lem64]{Lem64}
  C. E. Lemke and J. T. Howson,
  1964,
  "Equilibrium Points of Bimatrix Games",
  \textit{SIAM Journal on Applied Mathematics},
  12 (2), pp 413--423.

\bibitem[Nisan]{Nisan}
    Noam Nisan \textit{et al.},
    \textit{Algorithmic Game Theory},
    \url{https://www.cs.huji.ac.il/~noam/bn-ca.pdf}
    (accessed 6 August 2018).
    
\bibitem[Rap66]{Rap66}
  A. Rapoport and M. Guyer,
  1966,
  "A taxonomy of 2 x 2 games,"
  \textit{General Systems},
  11,
  no.~1,
  pp 203--214.

\bibitem[Rob05]{Rob05}
  David J. Goforth and David R. Robinson,
  \textit{Topology of the 2x2 Games},
  Routledge Advances in Game Theory,
  ISBN 9780415654586,
  CRC Press,
  2012.
  
\end{thebibliography}

%---------------------------------------------------------------
\end{document}